\documentstyle[12pt,twoside,fleqn,espcrc1,epsfig]{article}

\title{Space-time analysis: HBT at SPS and RHIC}

\author{Urs Achim Wiedemann\address{Physics Department, 
        Columbia University, \\ 
        10027 New York, U.S.A.}}

\begin{document}
\maketitle

\begin{abstract}
In this review talk, I first summarize the conclusions which
can be drawn from a particle interferometric analysis of CERN SPS
data and then I address related questions
which are relevant for the upcoming experiments at RHIC.
\end{abstract}

\section{Introduction}

The CERN SPS lead beam program has covered for the first 
time with high statistics all HBT radius parameters of the full 
three-dimensional parametrization over a wide region in rapidity 
and transverse momentum (for data, see Ref.~\cite{NA49}). Theory 
provides by now a detailed data analysis strategy and interpretation
(for a review, see Ref.~\cite{WH99}) . In this talk, I 
summarize the conclusions which we can draw from the recent
CERN SPS data, and I identify questions which can be
settled by the upcoming RHIC experiments. 

The main goal of HBT is to determine the geometry and dynamics of the
heavy ion collision at freeze-out. This information is becoming 
increasingly important for the discussion of other observables. For example: 
i) In studies of anisotropic flow~\cite{O98,PV98}, one aims at  
exploiting the HBT space-time picture to disentangle geometrical and 
dynamical components of anisotropy~\cite{VC96,W97,H98,TS99}. These cannot 
be separated 
on the basis of one-particle spectra alone. ii) In the search for event by
event multiplicity fluctuations of qualitatively new origin (e.g.
DCCs), fluctuations due to Bose-Einstein correlations are an important 
background source~\cite{P93}. Their calculation 
depends strongly on the phase-space density attained in the last stage 
of the collision. The HBT analysis of experimental data now provides 
estimates of this phase space density for a large variety of collision 
systems~\cite{FHTWC99}, and work on its consequences for 
multiplicity fluctuations is in progress~\cite{S99}. iii) 
In event generator studies, HBT measurements provide 
constraints on the simulated space-time structure~\cite{KKG}, and the 
question how microscopic dynamics can account for the observed final 
state emission region becomes of the utmost importance.

In the light of these rapidly developing applications, one should 
not forget the fundamental limitations of any particle interferometric
analysis. HBT extracts the phase space density $S(x,K)$ from a combination 
of hadronic one-particle and identical two-particle spectra of the abundant
hadron species:
\begin{eqnarray}
  E{dN\over d^3p} &=& \int d^4x\, S(x,p)\, ,
        \label{eq1}\\
  C({\bf K},{\bf q}) &=& 1 +
        {{ \vert \int d^4x\, e^{iq\cdot x}
            S(x,K)\vert^2 }
          \over
          {\int d^4x\, S(x,P_1)\, \int d^4y\, S(y,P_2)} }\, . 
        \label{eq2}
\end{eqnarray}
However: i) $S(x,K)$ characterizes the geometry and dynamics of the 
collision {\it at freeze-out} and thus provides only a snapshot of the 
last stage of the collision. What is characterized is the result of a
dynamical evolution, not the dynamical evolution itself. ii) HBT 
measures the relative distance between emission points. It does not 
measure the absolute position of emission points. This is of particular 
importance for the interpretation of the average emission time $\tau_0$, 
see below. iii) The measurement of $C(K,q)$ gives independent access to only 
three of the four space-time dependencies of $S(x,K)$. This is a 
consequence of detecting on-shell particles. It implies a model-dependent 
analysis strategy in which a model emission function $S(x,K)$ is compared to
data. 
%
\begin{center}
\begin{figure}[t]
\centerline{\epsfig{figure=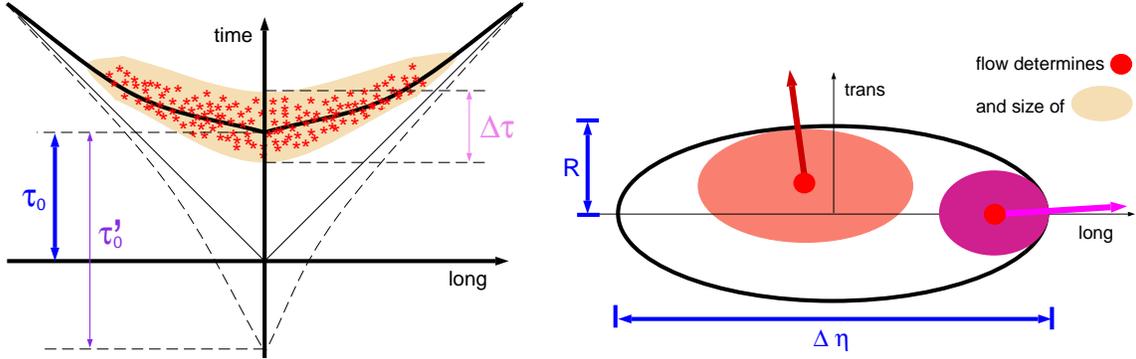,width=15.0cm}}
\vspace{-1.0cm}
\caption{Characterization of the pion emission function (\protect\ref{eq3})
in terms of Gaussian widths $R$, $\Delta\eta$, $\Delta\tau$. The strength
of the collective flow field $u_\mu$ determines the correlation between
points of highest emissivity and predominant emission direction. The
temperature $T$ determines the degree of thermal smearing around these
points of highest emissivity.
}\label{fig1}
\end{figure}
\end{center}

\vspace{-1cm}
\section{Final State Geometry and Dynamics of Pb+Pb at 158 GeV/c}

The expression for the one-particle and two-particle spectra
(\ref{eq1}) and (\ref{eq2}) in terms of the emission function
$S(x,K)$ form the starting point for 
various different approaches: i) In hydrodynamical simulations of heavy
ion collisions~\cite{B99,BRMG96}, they specify the 
integrations over the freeze-out hypersurface. ii) For event generator 
studies, a discretized formulation exists~\cite{KKG}. iii) 
Analytical model studies start from a
simple parametrization of the model emission function $S(x,K)$ whose
model parameters are then extracted from a fit of (\ref{eq1}) and 
(\ref{eq2}) to data. 
A simple, frequently used model ansatz
for the phase space distribution $S(x,K)$ in terms of very
few, physically intuitive fit parameters, is e.g. 
  \begin{eqnarray}
   S(x,P) &=& \textstyle{2J + 1 \over (2\pi)^3}\,
   P{\cdot}n(x)\,
   \exp{\left(- {P \cdot u(x) \over T} \right)}\,  
          \exp\left( - {r^2\over 2 R^2} 
                     - {\eta^2\over 2 (\Delta\eta)^2}
                     - {(\tau-\tau_0)^2 \over 2 (\Delta\tau)^2}
                 \right) \, . 
                 \label{eq3} 
  \end{eqnarray}
The physics contained in this ansatz is described elsewhere in 
much detail~\cite{WH99}. As shown in the sketch of Fig.~\ref{fig1},
the spatial extension of the pion emission
region is characterized by Gaussian widths $R$, $\Delta \eta$
and $\Delta\tau$ in the transverse, longitudinal and temporal
direction respectively. The absolute source position 
in the Minkowski diagram is fixed by the time scale $\tau_0$, and
the pion source emits particles at temperature $T$. The model
allows for a directed dynamical component via the flow field 
$u_\mu(x)$ which is assumed to show Bjorken 
scaling in the longitudinal direction superimposed with a transverse
flow component $\sinh \eta_t(r)$, $\eta_t(r) = \eta_f r/R$. To sum up: the
phase space density (\ref{eq3}) is characterized by only 6 
physically intuitive fit parameters: $T$, $\eta_f$, $R$, $\Delta\eta$, 
$\Delta\tau$ and $\tau_0$. 
%
\begin{center}
\begin{figure}[t]
\centerline{\epsfig{figure=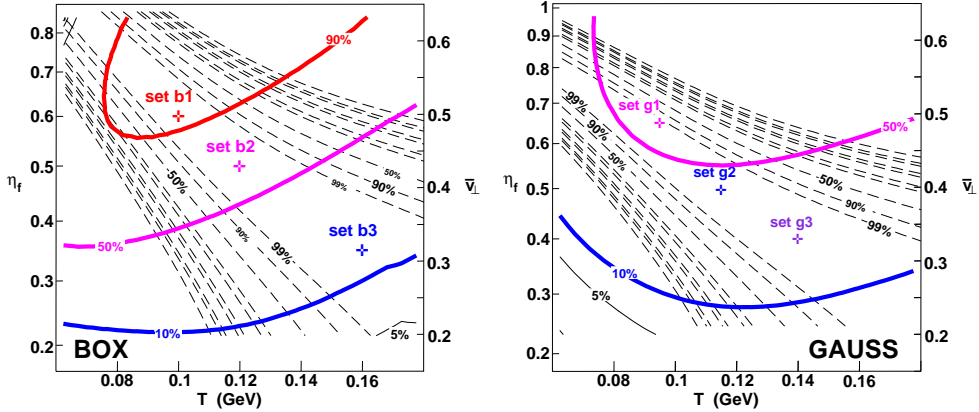,width=13.0cm}}
\vspace{-1.0cm}
\caption{Confidence levels for the optimal temperature $T$ and 
transverse flow $\eta_f$ values from fits of particle emission 
functions to NA49 one-particle spectra (dashed lines) and two-particle 
spectra (solid lines). The transverse density profile of model emission 
functions was chosen box-shaped (l.h.s.) and Gaussian (r.h.s.).
Figure from Ref.~\cite{T99}.
}\label{fig2}
\end{figure}
\end{center}
\vspace{-1.0cm}

Studies of many modifications of the parametrization (\ref{eq3}) 
exist. These include e.g. the effects of resonance decay 
contributions~\cite{WH96a}, other flow profiles~\cite{WSH96}, 
temperature gradients~\cite{CL96,TH97}, source opacity~\cite{HV96a} 
and azimuthal asymmetries~\cite{W97}. 
Comparisons of the particular model (\ref{eq3}) with experimental 
data exist for AGS~\cite{CN96} and SPS~\cite{WTH97,RNA49} 
energies. The 
main argument in favour of a largely model-independent interpretation
of the model parameters extracted in these studies is that
the measured particle spectra (\ref{eq1}) and (\ref{eq2}) are
mainly sensitive to the various r.m.s. of the emission function and
depend only weakly on the particular analytical implementation.
What matters is mainly the average transverse flow and the r.m.s. 
width of the transverse density profile, not their
particular $r$-dependence.
A recent analysis~\cite{T99} of NA49 data indicates the
limitations to these statements:

\begin{center}
\begin{figure}[h]
\centerline{\epsfig{figure=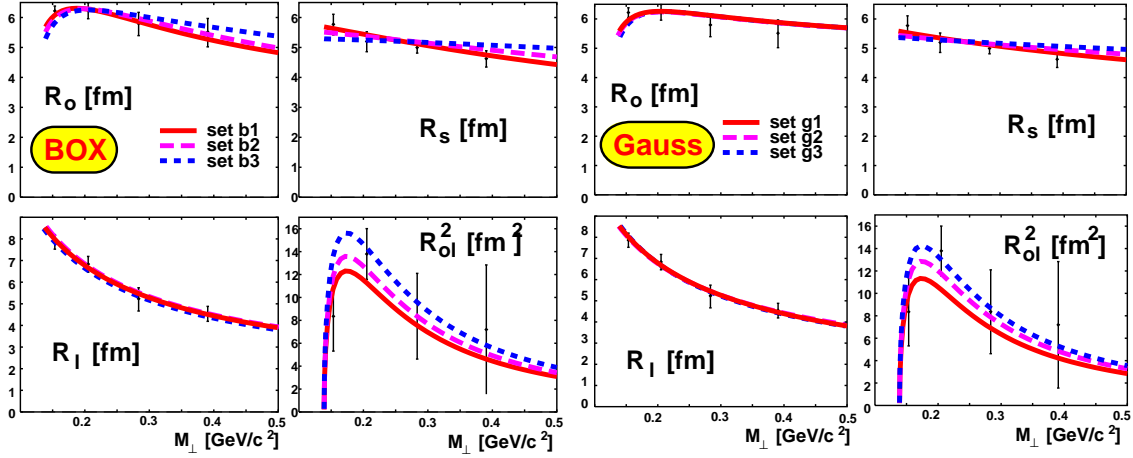,width=15.0cm}}
\vspace{-1.0cm}
\caption{Comparison of the NA49 Pb+Pb HBT radius parameters
at mid rapidity ($h^+\, h^+$ and $h^-\, h^-$ data are averaged)
to results from model emission functions with box-shaped (l.h.s.)
and Gaussian (r.h.s.) density profile. Different curves indicate
the parameter sets shown in Fig.~\protect\ref{fig2}. 
Figure taken from~\cite{T99}.
}\label{fig3}
\end{figure}
\end{center}
\vspace{-1.2cm}

B. Tom\'a\v{s}ik~\cite{T99} has compared the model emission function
(\ref{eq3}) to a model in which the Gaussian transverse density
distribution of (\ref{eq3}) was replaced by a box-shaped one,
$\exp\left( -{r^2\over 2\, R_G}\right) \longrightarrow \Theta(r-R_B)$.
For a comparison of the determined fit parameters of both models, one 
averages the transverse flow over the transverse density profile 
$H_{\rm profile}(r)$ and one recalls that the r.m.s. width for a 
Gaussian and a box profile differs by a factor 2:
\begin{eqnarray}
  \overline{v}_\perp &=& {1\over {\cal N}}\int r\, dr\, \tanh(\eta_t(r))\,
                       H_{\rm profile}(r)\, ,\\
  r_{\rm rms} &=& \sqrt{2}\, R_G = \textstyle{1\over \sqrt{2}} R_B\, ,
                 \qquad \eta_t(r) = \eta_f \textstyle{r\over r_{\rm rms}}\, .
\end{eqnarray}           
The slope of the transverse one-particle spectrum is due to a combination of
random (thermal) and directed (flow) transverse motion. Different
combinations of temperature $T$ and transverse flow $\eta_f$  
lead to the same effective blue-shifted temperature. As seen in 
Fig.~\ref{fig2}, the parameter combinations favoured by the fit
specify a sharp valley in the $T-\eta_f$-plane. The  
contour plots in this figure are similar for a Gaussian
and box-shaped density distribution, but they are not identical. 
The reason is that for an $r$-dependent flow field, particle emission
from different transverse distances $r$ occurs with different average
transverse momentum. Modifying the transverse density 
profile $H_{\rm profile}(r)$ can thus affect the $p_t$-dependence
of the transverse one-particle spectrum.
For the low-$p_t$ region ($p_t < 1.5$ GeV) measured by NA49,
these model-dependent variations are seen to be small. However, 
the high $p_t$ tails of the measured spectrum test in these analytical
models the large $r$-tails of the distribution $H_{\rm profile}(r)$
which is almost unconstrained by other observables. An application
of these models to higher $p_t$ (see e.g. Ref.~\cite{WA98}) may 
hence be expected to emphasize the model-specific $r$-dependence of
$H_{\rm profile}(r)$ and $\eta_t(r)$. The model-dependence increases
with $p_t$ - this is an important message for similar studies at RHIC.

HBT correlation radii allow to disentangle the $T-\eta_f$ ambiguity
of the fit to the one-particle spectrum. This is seen from the contour
plot of Fig.~\ref{fig2}. The contours for the 90, 50 and 10 \%
confidence levels are still very wide due to current experimental
uncertainties, but they clearly show a different $T-\eta_f$ correlation.
The fit favours a box-shaped density profile over a Gaussian one.
The parameter sets indicated by $b1$, $b2$, $b3$ and  $g1$, $g2$, $g3$
in Fig.~\ref{fig2} are used as input for the fits to the HBT radii in
Fig.~\ref{fig3}. In the $T-\eta_f$ plane, the fit is driven by the
slope of the side and out radius parameters, and the preferred values thus
depend strongly on taking all contributions to this slope properly
into account. All existing data analyses~\cite{CN96,WTH97,RNA49,T99} 
favour relatively low temperatures and large transverse flows.

The values~\cite{T99} of the model parameters used for the fits of 
Fig.~\ref{fig3} are summarized in the table at the end of this section.
The box-shaped and Gaussian model emission functions differ in several
details. Especially, a box-shaped density profile reproduces the data
with better $\chi^2$. A similar trend was also pointed out in a recent
study of deuteron coalescence models~\cite{SH99}. But irrespective of
these finer tendencies in the data, the main message is unaffected by
details of the analytical form of the emission function: we observe
a system with strong transverse flow which has expanded up to freeze-out
to twice its initial size, and which has cooled down substantially 
during this expansion.

It is an important problem to understand the dynamical origin of 
these source characteristics. One may e.g. question the dynamical
consistency of the presented data on the ground that the average
transverse velocity $\bar{v}_\perp$ and the emission time $\tau_0$
seem barely enough to account for an expansion to twice the initial
size. However, HBT does not measure absolute positions, and the
model parameter $\tau_0$ thus provides only a lower bound for the
total time after impact~\cite{WH99}. Also, very little is known
about the question how fast the final transverse flow of a 
hadronic system can be built up during the collision. At present,
one cannot even rule out the point of view that without novel
dynamical assumptions, none of the existing event generators can
account for the observed strong transverse expansion solely with
final state interactions~\cite{MG}. 

A constructive first step to clarify the dynamical origin of HBT 
fit parameters is the comparison of HBT radii from event generators
and experiment combined with a comparison of the 
simulated space-time structure to the model (\ref{eq3}). This may
help to understand how a dynamical model in which the time after
impact is known can account for the measured data.
%
\begin{center}
\renewcommand{\arraystretch}{1.15}
\begin{tabular}{|c|ccc|ccc|}
\multicolumn{7}{c}{ } \\[-1.3ex]
      \hline \hline
 & \multicolumn{3}{|c|}{box-shaped} & \multicolumn{3}{|c|}{Gaussian} \\ 
\cline{2-7}
\raisebox{1.5ex}{set} & 
{\sf b1} & {\sf b2} & {\sf b3} & {\sf g1} & {\sf g2} & {\sf g3}  \\
\hline
$T$ (MeV) & 100 & 120 & 160 & 100 & 120 & 160 \\
$\eta_f$ & 0.6 & 0.5 & 0.35 & 0.6 & 0.48 & 0.35 \\
 $R_B/R_G$ (fm) & $12.1\pm 0.2$ & $11.5\pm 0.2$ & $10.7\pm 0.2$ &
$6.5 \pm 0.1$ &$5.9 \pm 0.2$ & $5.6 \pm 0.1$  \\
$\tau_0$ (fm/$c$) &
$6.3\pm 1.1$ & $5.5\pm 1.1$ & $4.4\pm 3.5$ & 
$7.8 \pm 0.8$ & $6.6 \pm 0.9$ & $5.5 \pm 0.9$  \\
 $\Delta \tau$ (fm/$c$) &
$3.6\pm 0.6$ & $3.2\pm 0.7$ & $2.6 \pm 2.0$  &
$2.3 \pm 0.7$ & $2.3 \pm 0.7$ & $ 1.8 \pm 0.8$ \\
 $\Delta \eta$ (fixed) & 1.3 & 1.3 & 1.3 & 1.3 & 1.3 & 1.3 \\
\hline
 $\bar v_\perp$ & 0.5 & 0.43 & 0.33 & 0.46 & 0.39 & 0.29 \\
\hline\hline
\end{tabular}
\renewcommand{\arraystretch}{1}
\end{center}

\section{Phase space density}

Once the emission function $S(x,K)$ is determined, the particle
phase space density $f({\bf x}, {\bf p}, t)$ is obtained by 
integrating $S(x,K)$ over all particles emitted up to the time
$t$,~\cite{WH99}. Of particular interest is the spatially
averaged phase space density $\langle f\rangle({\bf K})$ which
is directly related to the measured particle 
spectra~\cite{B94,M96,FHTWC99}:
\begin{eqnarray}
   \langle f\rangle({\bf K}) = 
    { {\int d^3x\, f^2({\bf x},{\bf K}, t>t_f)}\over
    \int d^3x\, f({\bf x},{\bf K}, t>t_f)}
    = {dN\over dY\, M_\perp\, dM_\perp\, d\Phi}\,
    { \sqrt{\lambda(K_\perp,Y)}\over V_{\rm eff}(K_\perp,Y)}\, .
  \label{eq4}
\end{eqnarray}
This is a measure for the number of particles in a momentum bin 
$dY\, M_\perp\, dM_\perp\, d\Phi$ which are contained inside the
spatial volume $V_{\rm eff}(K_\perp,Y)$. The spatial volume
$V_{\rm eff}(K_\perp,Y)$ is determined
by the measured HBT radius parameters:
\begin{eqnarray}        
   V_{\rm eff}(K_\perp,Y) = {{M_\perp\, \cosh Y}\over \pi^{3/2}}\,
   \left( R_s\, \sqrt{R_0^2\, R_l^2 - (R_{ol}^2)^2}\right)\, .
   \label{eq5}
\end{eqnarray} 
The $\lambda$ intercept parameter in (\ref{eq4}) ensures that pions
from long-lived resonance decay contributions~\cite{WH96a}, which are emitted 
outside the collision region, do not contribute to the average phase 
space density $\langle f\rangle({\bf K})$.
%
\begin{center}
\begin{figure}[t]
\centerline{\epsfig{figure=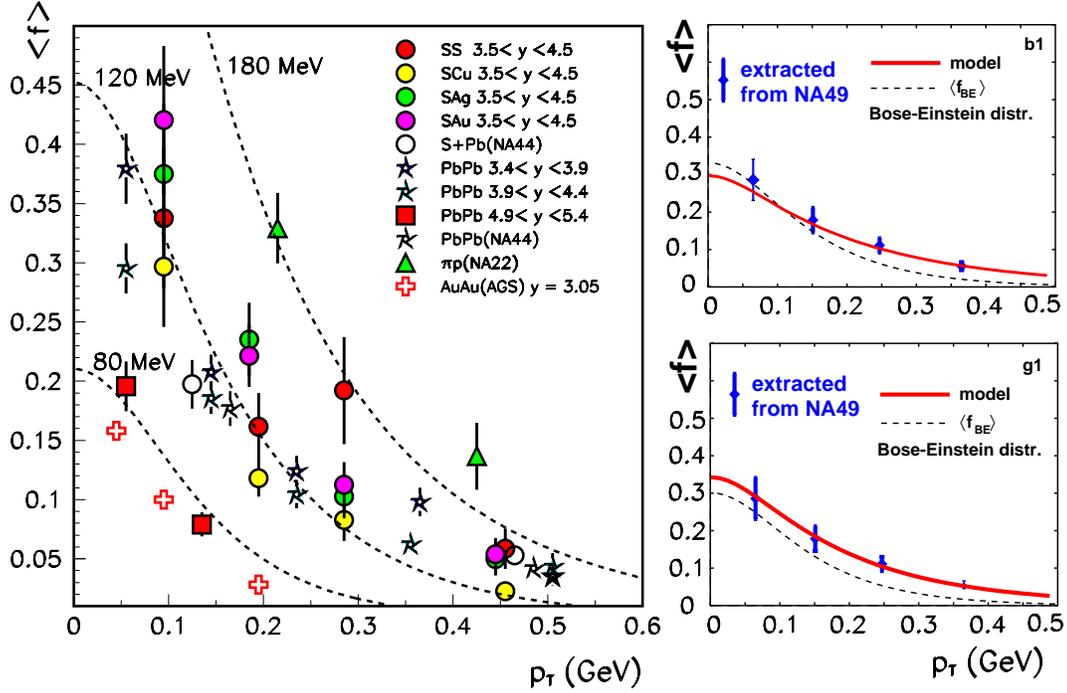,width=14.0cm}}
\vspace{-1.0cm}
\caption{Spatially average phase space density in narrow rapidity
windows as a function of the transverse momentum. Dashed lines
denote Bose-Einstein distributions at temperatures $T=$ 80, 120 and
180 MeV. Figure from ~\cite{FHTWC99} (l.h.s.) and ~\cite{T99} (r.h.s.).
}\label{fig4}
\end{figure}
\end{center}
\vspace{-0.9cm}

With the help of (\ref{eq4}), D. Miskowiec found in the E877 data
a factor 2 decrease of the phase space density 
in the very forward rapidity region. 
A comprehensive analysis~\cite{FHTWC99} of various different collision 
systems reveals now that such strong rapidity variations of 
$\langle f\rangle({\bf K})$ occur only very close to the 
kinematically allowed boundaries. As seen from Fig.~\ref{fig4}, 
except for the projectile rapidity region, $\langle f\rangle({\bf K})$ is
{\it almost universal}. The variation of $\langle f\rangle({\bf K})$ is
small compared to the order of magnitude change in the $dN^-/dY$ between 
the different collision systems (22 for S-S vs. 185 for Pb-Pb). 
The transverse momentum dependence of $\langle f\rangle({\bf K})$
is in rough agreement with a Bose-Einstein distribution. Deviations
from this show up in a weaker transverse momentum dependence of 
$\langle f\rangle({\bf K})$ and can be accounted for by transverse
flow. This is seen on the r.h.s. of Fig.~\ref{fig4}, where the average 
phase space density
extracted from NA49 mid rapidity data is compared to results from
the optimal model parameter sets 'b1' and 'g1' indicated in 
Fig.~\ref{fig2}. 

\section{Extrapolation of the AGS and SPS systematics to RHIC}
All observations indicate that the decoupling of pions occurs at an 
approximately constant phase space density irrespective of the 
collision system. This implies that the freeze-out volume scales 
with particle multiplicity
\begin{equation}
  V_{\rm eff}(K_\perp,Y) \propto {dN^-\over dY\, M_\perp\, 
  dM_\perp\, d\Phi}\, .
  \label{eq6}
\end{equation}
The qualitatively new aspect of Fig.~\ref{fig4} is
to establish  equation (\ref{eq6}) {\it differentially}, i.e. for 
narrow  bins in transverse momentum and rapidity. $K_\perp$-integrated
versions of (\ref{eq6}) are well-known. NA35 
observed e.g. for different collision systems 
a linear increase of the $K_\perp$-averaged collision volume 
with $dN^-/dY$ up to $dN^-/dY = 60$ ~\cite{NA35}. 
The $K_\perp$-averaged NA49 data for central rapidity follows the 
same systematics up to $dN^-/dY =180$, and RHIC will allow to test
this systematics even further.

Eq. (\ref{eq6}) can be used for extrapolation into this new high 
multiplicity regime accessible to RHIC:
The most naive RHIC-prediction is obtained by turning 
$V_{\rm eff} \propto dN/dY$ into a simple multiplicity dependence
for the side radius parameter which is typically the least affected 
by the dynamics, $R_s \propto \left(dN/dY\right)^{1/3}$. This leads to
$R_s \approx 9$ fm for ${dN^-\over dY} = 600$  and 
$R_s \approx 11$ fm for ${dN^-\over dY} = 1200$. Depending on the
dynamics, however, even if (\ref{eq6}) holds, the transverse radius 
parameters may be larger or smaller. One such example, based on
a hydrodynamical picture, assumes that the transverse flow gradient
does not change between heavy ion collisions at SPS and RHIC. From
this picture, one expects changes in the emission time $\tau_0$,
transverse radius $R$, and transverse flow $\eta_f$~\cite{H99}
        \begin{eqnarray}
            \tau_0\Bigg\vert_{\rm RHIC} \approx  
            2\, \tau_0\Bigg\vert_{\rm SPS}\, ,
            \qquad\quad
            R\Bigg\vert_{\rm RHIC} \approx  
            1.3\, R\Bigg\vert_{\rm SPS}\, ,
            \qquad\quad
            \eta_f\Bigg\vert_{\rm RHIC} \approx  
            1.3\, \eta_f\Bigg\vert_{\rm SPS}\, .
            \label{eq7}
        \end{eqnarray}
The resulting HBT radius parameters indicate a collision system which
is substantially more extended along the beam axis at the expense of
a relatively small increase in the transverse extensions,
\begin{eqnarray}
  R_s\vert^{\rm RHIC} \leq 1.2
  R_s\vert^{\rm SPS} < 8\quad \hbox{\rm fm}\, ,
  \qquad\quad
  R_\parallel\vert^{\rm RHIC} \approx 2 R_\parallel\vert^{\rm SPS}\, .
  \label{eq8}
\end{eqnarray}
Most scenarios favour a sudden bulk freeze-out of all pions which is
reflected in small values for $R_o^2-R_s^2$. However, hydrodynamical
scenarios with a particularly soft equation of state lead to large
differences between the two transverse parameters~\cite{RG96}: 
$R_o^2 \gg R_s^2$. In these scenarios, only
$\approx 50$ percent of the difference $R_o^2-R_s^2$ stems from
an extended lifetime of the system. An equally important contribution
to $R_o^2-R_s^2$ comes from a strong negative $x$-$t$ 
correlation~\cite{BRMG96}.

\section{Multiparticle Bose-Einstein symmetrization}
Standard calculations of the two-particle correlator $C(K,q)$ are
based on approximating $N$-particle symmetrized final state 
wavefunction by a sum of two-particle 
symmetrized contributions. For small phase space densities, this
is a valid approximation. However, if the
phase space density is sufficiently high, multiparticle 
symmetrization effects cannot be ignored. Recently, progress was made
in quantifying the decisive words "sufficiently high" of this last 
sentence in simple model studies: 

The well-known combinatorial problem is that one obtains from an  
$N$-particle symmetrized wavefunction $N!$ contributions
to the measurable particle spectra. Even for moderate event 
multiplicities, brute force numerical calculation do not work
($ 100! \approx 9*10^{157}$). However, for pion emitting
sources defined by the emission function $S(x,K)$ and the
event multiplicity $N$, Pratt's algorithm~\cite{P93} provides
a substantial simplification: 
%
\begin{center}
\begin{figure}[h]
\vspace{-1.0cm}
\centerline{\epsfig{figure=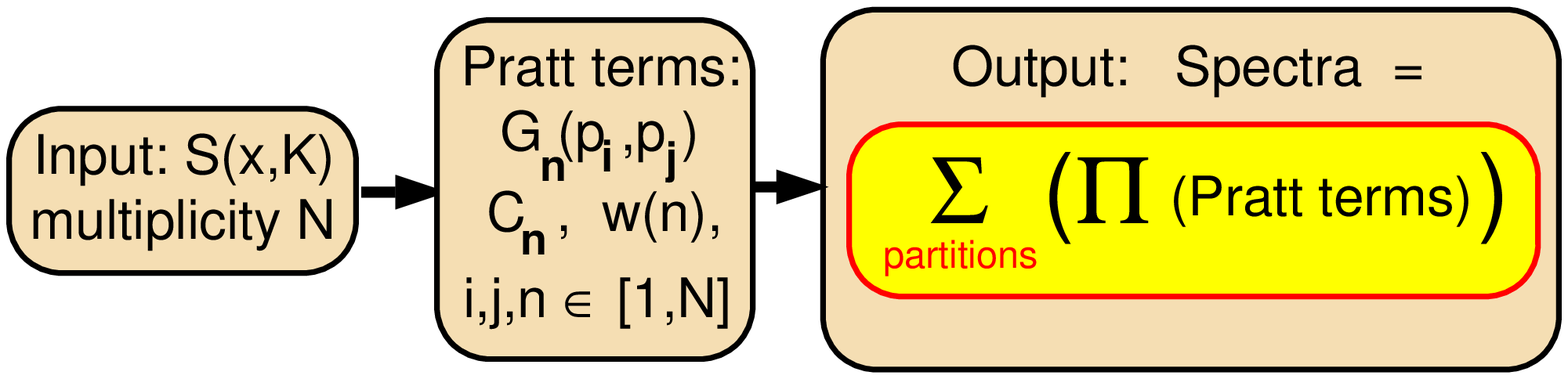,width=13.0cm}}
\vspace{-1.0cm}
\end{figure}
\end{center}

Here, the building blocks $G_n({\bf p}_i)$, 
$C_n$ and $w(n)$ are defined in terms of $n$-dimensional
integrals over $S(x,K)$, $n\leq N$. All measured spectra reduce
to sums over products of these building blocks. While the 
technicalities of this approach cannot be explained shortly
(see Ref.~\cite{WH99} for a review), the important message is
that the sums involved in the final answer run over the
number of all partitions of $N$ different contributions, rather
than over $N!$. Note that Partitions[N=100] =$2*10^9$, and this
makes a numerical calculation feasible.

\begin{center}
\begin{figure}[b]
\centerline{\epsfig{figure=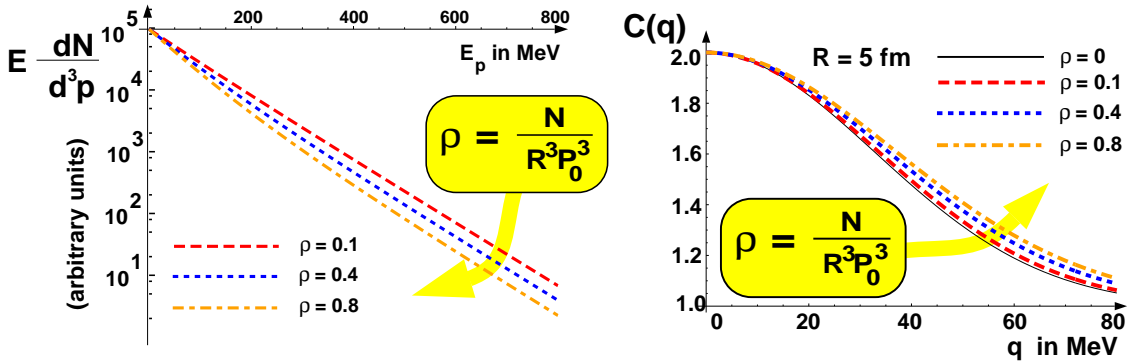,width=15.0cm}}
\vspace{-1.0cm}
\caption{Modifications of one-particle spectrum (l.h.s.) and
two-particle correlator (r.h.s.) due to multiparticle symmetrization
effects. With increasing phase space density, the one-particle
spectrum gets steeper and the two-particle spectrum gets wider.
Figure taken from~\cite{W98}.}
\label{fig5}
\end{figure}
\end{center}

The remaining computational task is the
calculation of the $n$-dimensional integrals $G_n$, $C_n$ and
$w(n)$. For Gaussian emission functions, this can be done 
analytically~\cite{CZ97,W98}. Fig.~\ref{fig5} shows the results
of such a model study and confirms the observations of Zajc~\cite{Z87}. 
Pions like to sit close together in 
momentum space and this leads to a steepening of the one-particle
spectrum. Also, they like to sit close together in configuration space
and hence the two-particle correlator gets wider. As
can be seen from Fig.~\ref{fig5}, these effects are small but 
visible for realistic phase space densities below $1\,\, \hbar^3$.
%
\vspace{-1cm}

However, existing calculations like Fig.~\ref{fig5} cannot be compared
to experiment. The problem is that despite some recent progress~\cite{B96}, 
there is no realistic description of multiparticle final state Coulomb 
interactions. The HBT effect and Coulomb repulsion occur both on
the same relative momentum scale and counteract each other. One may 
hence expect that their multiparticle contributions cancel to a large extent. 
Thus, present calculations can be expected to provide a loose upper bound 
on multiparticle symmetrization effects. The important message for
RHIC is that if the so far universal value of 
$\langle f\rangle({\bf K})$ is confirmed at RHIC, multiparticle
effects to the momentum spectra are of minor importance.

\section{Multiplicity fluctuations}

There is one important observable where the calculation of multiparticle 
symmetrization effects allows for a comparison with experimental data
and that is the modification of the charged to neutral multiplicity 
distribution due to Bose-Einstein statistics. A typical Bose-Einstein
modified distribution, proposed originally for the description of
anomalous cosmic ray (CENTAURO) events, is~\cite{PZ94}
\begin{eqnarray}
  P(n_0,n_+,n_-) = {1\over Z}\, {N!\over 3^N}\, 
    {w(n_0)\over n_0!}\, {w(n_+)\over n_+!}\,  {w(n_-)\over n_-!}\,
    \Bigg\vert_{N=\sum n_i}\, .
    \label{eq9} 
\end{eqnarray}
Here, the total number $N$ of neutral and charged pions is kept constant,
but one assumes that the multiplicity distribution is arranged according
to BE-statistics via the channel $\pi^0\, \pi^0$ 
$-$ $\pi^+\, \pi^-$. $Z$ is a normalization, and
$w(n)$ denotes the multiparticle symmetrization weights which include 
higher order correlations. According to Pratt's formalism, they are
given by a sum over all partitions $(m,l_m)_n$ of the building 
blocks $C_m$~\cite{P93}
\begin{eqnarray}
   w(n) = \sum_{(m,l_m)_n} 
        { C_1^{l_1}\,C_2^{l_2}\, \cdots\, C_m^{l_m}\over
          {\prod_m\, m^{l_m}\, (l_m!)}} 
   \label{eq10} 
\end{eqnarray}
These expressions simplify substantially for Gaussian models of the
emission function $S(x,K)$. One observes that the leading contribution
to the building blocks $C_m$ is a simple polynomical in the phase
space volume $\epsilon$~\cite{W98},
        \begin{eqnarray}
          C_m \approx \epsilon^{m-1}\, ,
          \qquad \hbox{where}\qquad \epsilon
          = {\hbar^3\over R^3\, P_0^3}\, ,
          \label{eq11}
        \end{eqnarray}
and this reduces the complicated combinatorics of (\ref{eq10}) to
a simple polynomial~\cite{W98}
        \begin{equation}
                w(n) = \prod_{k=1}^n \bigl( 1 + 
                \epsilon(k-1)\bigr)\, .
                \label{eq12}
        \end{equation}
Based on equations (\ref{eq9}) - (\ref{eq12}), P. Steinberg~\cite{S99} 
has studied the charged to neutral particle ratio measured by WA98.
He starts from the observation that the distribution of this ratio
is approximately 15 \% broader than that obtained in a VENUS 4.12
event generator simulation. Clearly, this discrepancy can have many
reasons. The strong working hypothesis in Ref.~\cite{S99} is that
the event generator describes the physics correctly but misses
multiparticle symmetrization effects (as all Monte Carlo models
do). As a remedy of the latter, the VENUS 4.12 multiplicity 
distribution was modified via (\ref{eq9}). P. Steinberg finds
that for phase space densities consistent with those shown in
Fig.~\ref{fig4}, multiparticle correlations can account for a
substantial part of the observed discrepancy (approx. 8-10 \%).
Irrespective of the real origin of the discrepancy, this is a
remarkable result: even for the relatively small measured
phase-space densities, Bose-Einstein statistics leads to visible
changes in multiplicity distributions. This also indicates that
Bose-Einstein correlations are an important background source
in searches of qualitatively new physics (e.g. DCCs) in multiplicity 
distributions. 

I thank U. Heinz, J.G. Cramer, D. Ferenc and B. Tom\'a\v{s}ik with
whom I had the privilege to collaborate. This work was sponsored by
DOE Contract No. De-FG-02-92ER-40764.

\end{document}